\documentstyle{proceeding}
\input psfig.tex

\begin{document}
\title{A search for distant X-ray galaxy clusters}
\author{D.J. Burke\inst{1}, C.A. Collins\inst{2}, R.C. Nichol\inst{3}, A.K. Romer\inst{4},
B. Holden\inst{3}, R.M. Sharples\inst{1} \and M.P. Ulmer\inst{4}.}
\institute{Department of Physics, University of Durham, South Road, Durham, DH1 3LE, UK
\and Astrophysics Group, Liverpool John Moores University, Byrom Street, Liverpool, L3 3AF, UK
\and Department of Astronomy and Astrophysics, University of Chicago, 5640 S. Ellis Rd, Chicago, IL 60637, USA
\and Deptartment of Physics and Astronomy, Northwestern University, 2145 Sheridan Road, Evanston, IL 60208, USA
}
\maketitle

\begin{abstract}
We report on the progress of our search for serendipitous distant
X-ray galaxy clusters in the ROSAT PSPC pointed observations archive.
The initial aim of our work is to measure the X-ray luminosity function
of clusters down to $f_{lim} = 4 \times 10^{-14}$ erg cm$^{-2}$ s$^{-1}$
and so test the claims of cluster evolution at moderate redshifts.
We have R band images and spectroscopy for 45 extended sources
detected in 50 of the 100 deepest southern fields.
A preliminary analysis suggests we cover a range of distances,
from nearby clusters to possible distant clusters, $z \simeq 1.0$.
\end{abstract}

\section{Introduction}
One of the major goals of cosmology is to probe the evolution of structure with
lookback time.
Galaxy clusters play a key role in this ambition because:
\begin{itemize}
\item Their formation by gravitational collapse is understood well enough
to allow comparisons of observations with theoretical
predictions of density evolution.
\item They can be observed out to cosmologically significant distances ($z \simeq 1.0$)
and so provide an important probe of the conditions in dense environments at early
times.
\end{itemize}

Optically selected distant cluster catalogues (e.g. Gunn et al.~1986; Couch et al.~1990)
are plagued by selection effects such as cluster mis-identification due to line
of sight projection effects (Frenk et al.~1990).
X-ray cluster selection greatly reduces such effects and produces cleaner,
statistically better defined samples (Henry~1992).

Two recent surveys measured the
X-ray luminosity function (XLF) and found that there are
fewer X-ray bright clusters at $z \simeq 0.1 - 0.3$ than locally
(Edge et al.~1990; Henry et al.~1992).
This result, if true, is of major consequence to our understanding of cluster
evolution (e.g. Kaiser~1991).
Unfortunately the conclusions are tentative because the surveys become seriously
incomplete at the redshifts where the evolution is seen.

The increased spatial resolution and sensitivity of the ROSAT PSPC detector
compared to previous X-ray imaging satellites and its low background rate mean
that we can measure the cluster XLF at fainter fluxes.
We can therefore test both the claims of cluster evolution and measure
the shape of the local XLF.
Several groups are working towards this goal (e.g. RIXOS, RDCS \& WARPS),
each using different cluster selection methods.
As yet, there is no agreed method for detecting clusters in X-ray
imaging data and so it is important for independent surveys to investigate
different selection algorithms.
A comparison of these surveys is given in Table~\ref{tbl:surveys}.
\begin{table}
\caption{A comparison of distant X-ray cluster surveys.}
\label{tbl:surveys}
\[
\begin{array}{p{0.3\linewidth}c@{\hspace{0.3in}}c} \hline 
\noalign{\smallskip} Name & f_{lim}^\dagger/10^{-14} $ erg cm$^{-2} $ s$^{-1} & \Omega/$deg$^{2} \\
\noalign{\smallskip} \hline
\noalign{\smallskip}
This Project & 4        & 14 \\
EMSS         & \sim 10 & 40\\
RDCS         & 1        & 26\ddagger \\
RIXOS        & 3        & 15\ddagger \\
WARPS        & 7        & 13\ddagger \\
\noalign{\smallskip} \hline
\end{array}
\]
\begin{list}{}{}
\item[$^{\rm \dagger}$] Flux limits are for the 0.5 - 2.0 keV band.
\item[$^{\rm \ddagger}$] Areal coverage as given in 
Rosati 1995, Castander et al.~1995 and Jones et al.~1995.
\end{list}
\end{table}

\section{Source Selection}
\label{sec:source-selec}

To combat the severe contamination expected from stars and AGN at our survey
depths (Stocke et al.~1991), our primary selection criterion is source
extension.
As clusters have significantly harder spectra than the general population of X-ray
sources (Ebeling~1993)
we use source hardness as a secondary selection criterion.
The analysis is restricted to the 0.5-2.0 keV band, to reduce the contamination
from soft sources, and to the central region of the PSPC detector, where
the PSF does not change significantly with off-axis angle.
We are concentrating on the 100 deepest pointings which satisfy:
\begin{itemize}
\item T $\geq$ 10ks
\item $\left| b \right| > 20^0$
\item $\delta < 20^0$
\end{itemize}

Analysis begins with screening out periods of bad aspect error or
high particle background.
We then compute a global estimate for the background and search 
for sources using the Cash statistic (Cash~1979).
Sources are tested for extent by comparing the photon distribution
to the PSF, taking care to model both the positional and spectral
dependence of the PSF.
Monte Carlo simulations show that we can reliably use source extent
as a discriminant within the central region ($r \leq 18'$)
and that we expect to be 90\% complete out to $z \simeq 0.5$.
All sources which have an extended profile are selected for optical
imaging and spectroscopy, during which hardness ratios are used to
help identify the X-ray source.

\section{Optical Follow Up}
\label{sec:optical}

We have analysed 50 ROSAT fields and our initial
observing run, using EFOSC on the ESO 3.6m, produced
R band images and spectroscopy for the extended sources in these fields.
These sources can be characterised as:
\begin{itemize}
\item Pairs of stars with small projected separation.
\item Nearby clusters with $z < 0.3$.
\item Imaging suggests a cluster, but
spectroscopic confirmation was not possible. We estimate,
from the brightest galaxy magnitudes, $z = 0.3 - 0.5$.
\item Sources with no obvious identification even after imaging
to $R \simeq 22$.
\end{itemize}

\begin{figure}
\psfig{figure=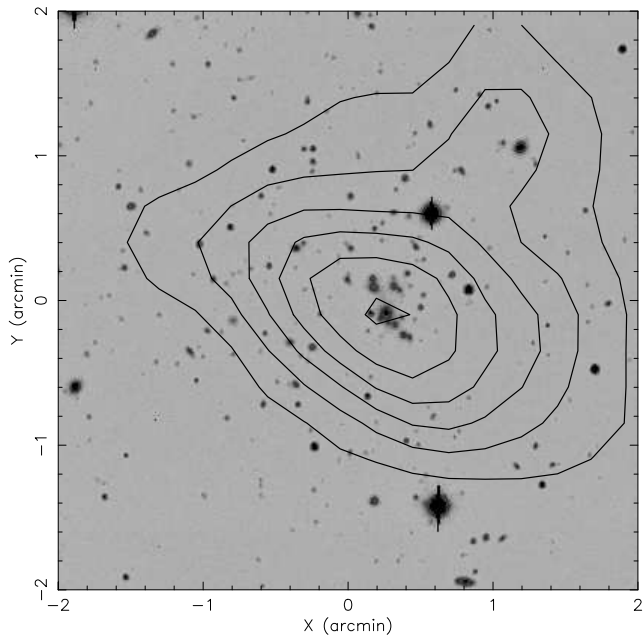,width=8.6truecm,angle=0}
\caption[]{X-ray contours (lightly smoothed) overlaid on R band image. At $z = 0.55$
1 arcmin equals 440$h_{50}^{-1}$ kpc ($q_0 = 0.5$).
\label{fig:cluster}
}
\end{figure}

We show, in fig~\ref{fig:cluster}, one of our distant cluster
candidates. Spectroscopy of the central galaxies suggest $z = 0.55$.
Fitting the X-ray spectrum by a Raymond-Smith plasma code, with
T=6 keV and half solar metallicity,
gives L$_X \sim 3 \times 10^{44}$ erg s$^{-1}$ (0.5-2.0keV).

\section{Future Work}
\label{sec:future}

The initial goal of this work is to measure the cluster XLF.
We have time allocated to complete the identification and spectroscopy
of our remaining fields.
Time has been applied for to secure identification and redshifs
of our unconfirmed cluster candidates.
As we suspect our unidentified sources may well be distant clusters,
we have applied for K band imaging.
This will allow us to identify clusters and we can use the K-$z$
diagram (Collins \& Mann 1995) to estimate cluster redshifts
with sufficient accuracy for construction of the cluster XLF.

Our collaboration is also using a wavelet-based detection algorithm
to search for clusters in the deepest Northern
ROSAT pointings.
Optical
follow up of these extended sources has begun on the ARC 3.5m telescope
at Apache point.
We will therefore be able to use our two well defined
catalogues to directly compare selection techniques and
assess our completeness.

Future work will investigate the relationship between the hot intracluster
gas and the optical properties of clusters and to study the Butcher-Oemler
effect (Butcher \& Oemler~1984).
Our sample, combined with other distant cluster catalogues
currently being compiled (e.g. RIXOS, RDCS and WARPS), will make an
excellent target list for AXAF and the upcoming 8m class of ground based
telescopes.

\begin{acknowledgements}
This research has made use of data obtained from the Leicester Database
and Archive Service at the Department of Physics and Astronomy, Leicester
University, UK.
\end{acknowledgements}

\end{document}